\documentclass[10pt]{article}
\pdfoutput=1
\usepackage{amssymb,amsmath,enumerate}
\usepackage{graphicx,rotate,multicol}
\usepackage{wrapfig}
\usepackage[colorlinks=true,
            linkcolor=red,
            urlcolor=blue,
            citecolor=blue]{hyperref}
\usepackage{color}
\usepackage{soul}
\usepackage{cite}
\long\def\rpl#1!!#2!!{\textcolor{red}{#1} \textcolor{blue}{#2}}

\def\tb{\tan\beta}
\def\cba{\cos(\beta-\alpha)}

\def \order(#1){{\cal O} \left(#1 \right)}

\evensidemargin=\oddsidemargin
\def\Eqn#1{Eq.\ (\ref{#1})}

\allowdisplaybreaks

\begin{document}
\hspace*{\fill}FTUV-17-1130\\
\hspace*{\fill}IFIC-17/55

\begin{center}
	{\Large \bf Can measurements of 2HDM parameters provide hints for
	\\ high scale supersymmetry? } \\
	\vspace*{1cm} {\sf Gautam
          Bhattacharyya$^{a,}$\footnote{gautam.bhattacharyya@saha.ac.in},~Dipankar
          Das$^{b,c,}$\footnote{dipankar.das@uv.es},~M. Jay
          P\'{e}rez$^{b,e,}$\footnote{mperez75@valenciacollege.edu},~Ipsita
          Saha$^{d,}$\footnote{ipsita.saha@roma1.infn.it},~Arcadi
          Santamaria$^{b,}$\footnote{arcadi.santamaria@uv.es}, \\~Oscar
          Vives$^{b,}$\footnote{oscar.vives@uv.es}} \\
	\vspace{10pt} {\small \em $^a$Saha Institute of Nuclear
          Physics, HBNI, 1/AF Bidhan Nagar, Kolkata 700064, India \\
          $^b$Departament de F\'{i}sica T\`{e}orica,
          Universitat de Val\`{e}ncia and IFIC, Universitat de
          Val\`{e}ncia-CSIC, \\ Dr. Moliner 50, E-46100 Burjassot
          (Val\`{e}ncia), Spain \\
          $^c$Department of Physics, University of Calcutta, 92
          Acharya Prafulla Chandra Road, Kolkata 700009, India \\
		$^d$Istituto Nazionale di Fisica Nucleare, Sezione di Roma,
		P. le Aldo Moro, 2, 00185, Roma, Italy \\
                $^e$Valencia State College, Osceola Campus, 1800 Denn
                John Ln, Kissimmee, FL, USA}
	
	\normalsize
\end{center}

\begin{abstract}

Two-Higgs-doublet models (2HDMs) are minimal extensions of the
Standard Model (SM) that may still be discovered at the LHC.  The quartic
couplings of their potentials can be determined from the measurement of
the masses and branching ratios of their extended scalar sectors.  
We show that the evolution of these couplings through renormalization 
group equations can determine whether the observed 2HDM is a low energy 
manifestation of a more fundamental theory, as for instance, supersymmetry, 
which fixes the quartic couplings in terms of the gauge couplings.
At leading order, the minimal supersymmetric extension of the SM (MSSM) dictates all the quartic couplings, which can be translated into a predictive structure for the scalar masses and mixings at the weak scale.
Running these couplings to higher scales, one can check if they
converge to their MSSM values, and more interestingly, whether one can
infer the supersymmetry breaking scale. Although we study this question in the
context of supersymmetry, this strategy could be applied to any theory whose 
ultraviolet completion unambiguously predicts all scalar quartic couplings.
\end{abstract}

\bigskip

\section{Introduction} \label{sec:intro}
Despite several theoretical and experimental motivations for new
physics beyond the standard model~(BSM), the LHC data on the
production and branching ratios of the Higgs boson are tantalizingly
close to their SM predictions
\cite{Chatrchyan:2012xdj,Aad:2012tfa,Khachatryan:2014jba,Khachatryan:2016vau},
yet to reveal any convincing sign of life beyond it.  Although it is
entirely possible that no new physics exists between the electroweak
and the grand unification or Planck scales, {\em modulo} some dark
matter source, this \emph{grand desert} scenario carries with it many
unpleasant features such as the hierarchy between the electroweak and
Planck scale, lack of gauge unification, etc.

From a rather pragmatic point of view, a vital question in motivating
collider searches for BSM physics above the electroweak scale (EWS) is
whether we can foretell the approximate mass range where new particles
are expected to appear. This is precisely the hurdle most new physics
models fail to cross. Predicting the existence of many new particles,
most such models provide guidance on \emph{what} to look for, leaving
no clue, unfortunately, on \emph{where}. Therefore, any principle that
gives some idea of the probable mass scales of the new particles
deserves special attention from a phenomenological point of view. The
present article is an attempt in this direction.

Amongst the vast array of BSM scenarios, extensions of the SM scalar
sector have been explored with various motivations, e.g.~for
generating the necessary additional CP violation to account for the
observed baryon asymmetry of the universe. In these models, the
125~GeV Higgs boson just happens to be one of the scalars in the spectrum, the others 
yet to be discovered. Phenomenologically, 2HDMs
provide one of the simplest realizations in this category, wherein the
scalar sector of the SM is augmented by just one additional doublet
\cite{Branco:2011iw,Bhattacharyya:2015nca}. Aside from their
simplicity, these models have the desirable property that the oblique
electroweak $\rho$-parameter remains unity at the tree level, along
with providing an \emph{alignment limit}
\cite{Gunion:2002zf,Carena:2013ooa,Dev:2014yca,Bhattacharyya:2014oka,Das:2015mwa},
where a SM-like Higgs can be recovered. Quite notably, the MSSM~\cite{Nilles:1983ge,Haber:1984rc,Lykken:1996xt,Martin:1997ns,Drees:2004jm,Baer:2006rs}
is structured around two such Higgs doublets.

In view of the increasing affinity of the LHC Higgs data to the
SM-predicted values, the ability to attain the alignment limit might
hold the key for future survival of any such new physics model.  Let
us suppose that the only hint of new physics from forthcoming data
somehow points towards a 2HDM structure, either directly or
indirectly. If the 2HDM is viewed as an {\em effective low-energy}
model arising from a more fundamental ultraviolet (UV) theory, it would be
interesting to ask whether the knowledge of the 2HDM potential at LHC
scales can give us any hint of its embedding in a particular UV
scenario, containing massive states sitting at an inaccessibly high
scale.
   
Under the assumption of CP conservation, 2HDMs predict the existence
of five physical scalars: two CP-even ($h$ and $H$), one CP-odd~($A$)
and a pair of charged scalars~($H^\pm$).  We assume that the lightest
CP-even scalar~($h$) is the one already discovered with a mass $\sim
125$~GeV. Its SM-like properties, as LHC data indicate, compel us to
stay close to the alignment limit. However, it is still possible that
the nonstandard scalars might all be lurking below the TeV scale
waiting to be discovered at the LHC. In that case, we would, in
principle, be able to measure all the parameters of the 2HDM scalar
potential.  By studying the renormalization group (RG) evolution of
these parameters, we would then be able to test whether the 2HDM is a low
energy manifestation of a more fundamental theory with an enhanced
symmetry at a high scale, $\Lambda_S$.

In this paper, we focus our attention on the MSSM framework as a well
motivated example in this category.  A high supersymmetry (SUSY)
breaking scale may seem to run contrary to the common lore of its
solution to the hierarchy problem and arriving at the correct $125$
GeV mass for the light Higgs. Nevertheless, viewed as a 2HDM effective
theory, achieving the correct mass for the Higgs can be translated
into obtaining the correct value for its quartic couplings when
matched and run down from $\Lambda_S$ to the EWS. Indeed, such ``High
Scale SUSY" scenarios have been studied before in the literature, both
in the case where the effective theory below the SUSY scale is
strictly the SM
\cite{Giudice:2011cg,Arvanitaki:2012ps,Bagnaschi:2014rsa,Vega:2015fna,Isidori:2017hac},
or a 2HDM in the context of a moderately high SUSY scale $ \Lambda_S
\sim 10^4$ GeV
\cite{Carena:2015uoe,Athron:2016fuq,Staub:2017jnp,Haber:2017erd,Chalons:2017wnz}. It
was found that, in both scenarios, solutions indeed exist for low
values of $\tan \beta$. Larger values of $\Lambda_S$ have also been
considered in \cite{Lee:2015uza,
  Bagnaschi:2015pwa,Ellis:2017erg,Wells:2017vla} where the 2HDM
spectrum was obtained by matching the 2HDM to the MSSM at the
$\Lambda_S$ scale and running it down to the EWS. These studies have
been done using state of the art calculations (matching at one loop
plus dominant two loops, and running at two loops). On the contrary,
here we follow a bottom-up approach by assuming that the spectrum of
scalar masses and mixings will be measured at the EWS from where the
scalar potential can be determined. Then, we run the quartic couplings
of the potential, using the 2HDM Renormalization Group Equations
(RGE)~\cite{Machacek:1983fi,Machacek:1983tz,Machacek:1984zw} as
implemented by SARAH 4~\cite{Staub:2013tta}, and check if they satisfy
the SUSY boundary conditions at a higher scale, as usually done in
Grand Unified Theories. This approach has the advantage that it is
independent of the details of the underlying theory which are hidden
in the matching conditions at the high scale.


\section{Effective 2HDMs and parameter counting} \label{sec:counting} 
The most general gauge invariant potential built with two SU(2)
doublet scalars (with hypercharge $Y=+1$), $\phi_1$ and $\phi_2$,
can be written as\cite{Gunion:2002zf}
%
\begin{eqnarray}
 V &=& m_{11}^2 \phi_1^\dagger\phi_1 + m_{22}^2\phi_2^\dagger\phi_2
 -\left(m_{12}^2 \phi_1^\dagger\phi_2 +{\rm h.c.} \right) \nonumber
 \\ & + &\frac{\lambda_1}{2} \left(\phi_1^\dagger\phi_1\right)^2
 +\frac{\lambda_2}{2} \left(\phi_2^\dagger\phi_2 \right)^2 +\lambda_3
 \left(\phi_1^\dagger\phi_1 \right) \left(\phi_2^\dagger\phi_2 \right)
 +\lambda_4 \left(\phi_1^\dagger\phi_2 \right)
 \left(\phi_2^\dagger\phi_1 \right) \nonumber \\ & +
 &\left[\frac{\lambda_5}{2}\left(\phi_1^\dagger\phi_2 \right)^2
 +\left(\lambda_6\left(\phi_1^\dagger\phi_1 \right) + \lambda_7
 \left(\phi_2^\dagger\phi_2 \right)\right)\left(\phi_1^\dagger\phi_2
 \right) +{\rm h.c.}\right] \,.
\label{potential}
\end{eqnarray}
This potential contains 3 mass parameters and 7 quartic couplings,
understood as $\overline{\rm MS}$ parameters at the
EWS arising from a more complete theory at higher energies.  If
$m_{12},\lambda_5,\lambda_6,\lambda_7$ are all zero, the potential has
an additional U(1) global symmetry~\cite{Bhattacharyya:2013rya}. If only
$m_{12},\lambda_6,\lambda_7$ are zero the U(1) is broken but there
remains an unbroken discrete $Z_2$ symmetry. If only
$\lambda_6,\lambda_7$ are zero, then this discrete symmetry is softly
broken by $m_{12}$. Models in which this $Z_2$ is also preserved by
the Yukawa couplings (with $\phi_1$ coupled only to down-type fermions
and $\phi_2$ only to up-type fermions) are called type II 2HDMs.  This
discrete symmetry is useful for avoiding large flavor changing neutral
currents and appears, as an approximate symmetry, in supersymmetric
models.

We assume that all the parameters in the potential are real. After electroweak 
spontaneous symmetry breaking (SSB), one obtains the
physical spectrum, specified by seven parameters: the four
physical scalar masses ($m_h$, $m_H$, $m_A$ and $m_+$), the total
vacuum expectation value~(vev) $v=\sqrt{v_1^2+v_2^2}$,
$\tan\beta=v_2/v_1$, and the alignment angle $\cos(\beta-\alpha)$
(here $\alpha$ is the mixing angle in the CP-even sector).

In principle, the whole spectrum can be determined from knowledge
of the quartic couplings. Consider the situation where all the
quartic couplings in \Eqn{potential} are known (from some symmetry
principle, e.g. supersymmetry) at a scale $\Lambda_S$. Then the
remaining three bilinear parameters can be solved from the knowledge
of $v$~($=246$~GeV), $m_h$~($\simeq 125$~GeV) and $\tan \beta$ (or
alternatively $\cos(\beta-\alpha)$). In other words, the complete 2HDM
spectrum is then determined {\em modulo} the experimental uncertainties
in the quoted parameters.  Explicitly, the SSB contributions to the
charged scalar masses and to the mass matrix of the neutral scalars in
the Higgs basis can be written solely in terms of the $\lambda_i$ and
$\tan\beta$, as follows (see \cite{Lee:2015uza,Bernon:2015qea} for details):
\begin{subequations}
\begin{eqnarray}
  g_{11}& =& \lambda_1 \cos^4 \beta + \lambda_2 \sin^4 \beta + 2
  \left(\lambda_3 + \lambda_4+\lambda_5\right) \sin^2 \beta \cos^2
  \beta + 4\lambda_6 \cos^3\beta \sin\beta + 4\lambda_7 \sin^3\beta
  \cos\beta \,, \label{g11}\\ g_{12}& =& \cos \beta \sin \beta
  \left(\lambda_2 \sin^2 \beta - \lambda_1 \cos^2 \beta +
  \left(\lambda_3 + \lambda_4+\lambda_5\right) \cos 2\beta
  \right)\nonumber\\ && +  3(\lambda_7-\lambda_6)\sin^2\beta\cos^2\beta
  + \lambda_6 \cos^4\beta - \lambda_7
  \sin^4\beta\,, \label{g12}\\ g_{22}& =& \left(\lambda_1 +
  \lambda_2\right) \cos^2 \beta \sin^2 \beta - 2 \left(\lambda_3 +
  \lambda_4\right) \cos^2 \beta \sin^2 \beta\nonumber\\ && + \lambda_5
  (\sin^4\beta+\cos^4\beta)+(\lambda_7-\lambda_6)\sin 2\beta \cos
  2\beta \,, \label{g22} \\ g_{+}& =&
  \frac{1}{2}\left(\lambda_5-\lambda_4\right) \label{gp}\, .
\label{gs}
\end{eqnarray}
\end{subequations}
The diagonalization of the mass terms leads to 
\begin{subequations}
\begin{eqnarray}
  g_{11} v^2&=& m_H^2\cos^2(\beta -\alpha) + m_h^2 \sin^2(\beta -\alpha) \,,  \\
  g_{22} v^2&=& m_H^2\sin^2(\beta -\alpha) + m_h^2 \cos^2(\beta -\alpha) - m_A^2    \,,  \\
  g_{12} v^2&=& \left( m_{h}^2 - m_{H}^2\right) \cos(\beta -\alpha)\sin (\beta -\alpha) \,,  \\
  g_+ v^2& =& m_+^2 - m_A^2\, ,
\end{eqnarray}
\label{gij+}
\end{subequations}
which, when inverted, yield (in terms of the known $m_h$ and $v$)
\begin{subequations}
\begin{eqnarray}
  m_{H}^2& =& g_{11} v^2 +\frac{\left(g_{12} v^2\right)^2}{g_{11} v^2-m^2_h} \,,  \\
  m_{A}^2& =&  m_h^2 - g_{22} v^2 + \frac{\left(g_{12} v^2\right)^2}{g_{11} v^2-m^2_h}   \,,  \\
  m_{+}^2& =&  m_h^2 - g_{22} v^2 + \frac{\left(g_{12} v^2\right)^2}{g_{11} v^2-m^2_h} + g_+ v^2\,,  \\
  \cos\left(\beta - \alpha\right)& =& -~\mathrm{sgn}(g_{12})\bigg/\sqrt{ 1 + \left( \frac{g_{12} v^2}{g_{11} v^2-m^2_h} \right)^2} \, .
\end{eqnarray}
\label{mHA+}
\end{subequations}
The above equations explicitly show how the scalar masses and mixings
can be obtained, once all $\lambda_i$ are known, in terms of $v$,
$m_h$ and $\tan\beta$.

Assuming that all supersymmetric particles are much heavier than the
EWS, the MSSM provides a perfect example of a model where the Higgs
sector is a 2HDM. In this case, the Higgs quartic couplings come from
the supersymmetric $D$-terms and, at tree level, are simple functions
of the gauge couplings~\cite{Haber:1984rc,Chung:2003fi}:
\begin{eqnarray}
\lambda_1 = \lambda_2 = \frac{1}{4}\left(g^2+g_Y^{2}\right) \,, ~~
\lambda_3 =\frac{1}{4}\left(g^2-g_Y^{2}\right) \,, ~~
\lambda_4 = -\frac{g^2}{2} \,, ~~ \lambda_5=\lambda_6=\lambda_7=0 \,,
\label{SUSY}
\end{eqnarray}
where $g$ and $g_Y$ are the ${\rm SU(2)_W}$ and ${\rm U(1)_Y}$ gauge
couplings, respectively. Note that all mass terms are also generated
at tree level. In particular, the $m_{12}$ term, which breaks the
discrete $Z_2$ symmetry softly, is related to the bilinear $B\mu$
term in the SUSY potential. Therefore, at tree level, the MSSM leads to a
type~II 2HDM.  The relations of \Eqn{SUSY} should be understood to
hold at a scale $\Lambda_S$, where the general 2HDM is matched to the
MSSM. Below $\Lambda_S$, the RG evolution of the 2HDM parameters
should be used to obtain the potential at the EWS.  Since the boundary
condition $\lambda_5=\lambda_6=\lambda_7=0$ increases the symmetry of
the quartic part of the Lagrangian, these couplings will not be
generated by the RG evolution and will still be zero at lower
energies.

This simple picture is perturbed if we consider the higher-order
corrections to the Higgs potential.  In the MSSM, the $Z_2$ symmetry
is broken by the $\mu$-term in the superpotential ($\mu$ being the
Higgsino mass parameter), and this breaking affects the higher order
matching of all the $\lambda_i$ at the scale $\Lambda_S$. In
particular, $\lambda_{5,6,7}(\Lambda_S)$ will arise at higher loops but
will always be proportional, at least, to $\mu/\Lambda_S$
\cite{Haber:1993an}, which we will consider to be small. Then, as RG
evolution cannot generate them, it is reasonable to assume
$\lambda_5\simeq \lambda_6\simeq \lambda_7\simeq 0$.

Similarly, the effective couplings $\lambda_i (i=1,\dots,4)$ in
\Eqn{SUSY} receive corrections at the scale $\Lambda_S$ that depend on
the full supersymmetric spectrum. These corrections are, however,
sub-leading. In fact, the corrections proportional to the large third
generation Yukawa couplings come with a factor of $\mu/\Lambda_S$ or
$A_{t,b}/\Lambda_S$, $A_{t,b}$ being the trilinear soft-breaking
terms. In the following, we assume that $\mu/\Lambda_S,
A_{t,b}/\Lambda_S \ll 1$, and thus these corrections as well as the
smaller gauge corrections can be safely neglected. 

Under these assumptions, we have only four quartic couplings, which
can be determined from the scalar masses and mixings by inverting
Eqs.~(\ref{g11})--(\ref{gp}) and using \Eqn{gij+} as follows,
\begin{subequations}
\begin{eqnarray}
  \lambda_1&=& g_{11} + g_{22} \tan^2 \beta - 2 g_{12} \tan \beta  \,,  \\
  \lambda_2&=& g_{11} + g_{22} \cot^2 \beta + 2 g_{12} \cot \beta \,,  \\
  \lambda_3&=& g_{11} - g_{22} + 2 g_{12} \cot\left( 2 \beta\right) + 2 g_+\,,  \\
  \lambda_4& =&  - 2 g_+\,.
\end{eqnarray}
\label{lambda1234}
\end{subequations}
Once these couplings are determined at the EWS, including appropriate
radiative corrections\cite{Chalons:2017wnz,Braathen:2017jvs}, we can use the 2HDM RGE
to check whether their values correspond to the MSSM boundary
conditions at a high scale.

\section{RG analysis and pointers to high scale SUSY} \label{sec:rg}
To obtain a qualitative understanding of the RG evolution, we can
begin by simply using the one loop RGE, checking the stability of
these results under higher order corrections {\it a posteriori}.  At
one loop, the RG evolution of the gauge couplings is very simple and
can be easily integrated. We will be interested here in the
combination $(g^2 + g_Y^2)/4$ which, in a supersymmetric framework,
would fix the boundary values for $\lambda_1$ and $\lambda_2$.
The RG evolution of this combination at one loop is given by, 
\begin{eqnarray}
{\cal D}(g^2 + g_Y^2)  = \frac{ -3 g^4 + 7 g_Y^4}{8 \pi^2}\,,
\end{eqnarray}
where ${\cal D} \equiv d / d(\log M)$. Substituting their EWS values, we
obtain $\left.(-3 g^4 + 7 g_Y^4)/(8 \pi^2)\right|_{M_z} \simeq 0.003$,
{\it i.e.} this combination remains essentially constant at one loop.
On the other hand, the one loop RGE for the quartic couplings depend
on the gauge as well as Yukawa couplings as follows~\cite{Branco:2011iw}:
\begin{subequations}
\begin{eqnarray}
  {\cal D}\lambda_1 &=& \frac{1}{16 \pi^2} \bigg[\frac{3}{4} \left(3 g^4 + g_Y^4+ 2 g^2 g_Y^2\right) - 3 \lambda_1 \left(3 g^2 + g_Y^2\right) \nonumber \\ 
     && + 12 \lambda_1^2 + 4 \lambda_3^2 + 4 \lambda_3 \lambda_4 + 2 \lambda_4^2 +   4 \lambda_1 \left( 3 y_b^2 + y_\tau^2\right) - 12 y_b^4 -4 y_\tau^4 \bigg]\,, \\
  {\cal D}\lambda_2 &=& \frac{1}{16 \pi^2} \bigg[\frac{3}{4} \left(3 g^4 + g_Y^4+ 2 g^2 g_Y^2\right) - 3 \lambda_2 \left(3 g^2 + g_Y^2\right) \nonumber \\ 
     && +12 \lambda_2^2 + 4 \lambda_3^2 + 4 \lambda_3 \lambda_4
  +2 \lambda_4^2 + 12 \lambda_2 y_t^2 - 12 y_t^4 \bigg] \,,\\
  {\cal D} \lambda_3 &=& \frac{1}{16 \pi^2} \bigg [\frac{3}{4} \left(3 g^4 + g_Y^4- 2 g^2 g_Y^2\right) - 3 \lambda_3 \left(3 g^2 + g_Y^2\right)  \nonumber \\ && 
  +2 \left(\lambda_1 + \lambda_2\right) \left(3\lambda_3 + \lambda_4\right) + 4 \lambda_3^2 + 2 \lambda_4^2
  + 2 \lambda_3 \left( 3 y_t^2 + 3 y_b^2 + y_\tau^2\right)- 12 y_t^2y_b^2 \bigg] \,,\\
    {\cal D} \lambda_4 &=& \frac{1}{16 \pi^2} \bigg [3 g^2 g_Y^2 - 3 \lambda_4 \left(3 g^2 + g_Y^2\right) 
    \nonumber \\ 
       && +2\left(\lambda_1 + \lambda_2 + 4 \lambda_3 \right) \lambda_4 + 4 \lambda_4^2 + 2 \lambda_4 \left( 3 y_t^2 + 3 y_b^2 + y_\tau^2\right)+ 12 y_t^2y_b^2 \bigg ] \,,
\end{eqnarray}
\end{subequations}
where, $y_f$ stands for the Yukawa coupling of the fermion $f
\,(=t,b,\tau)$.  From these equations we see that only $\lambda_2$
should have significant evolution due to the large top Yukawa
coupling, $y_t \sim \order(m_t/(v\sin\beta))$.  This is true for
$\tan\beta \sim $~1--3, which is the relevant range for high scale
SUSY, as we will see below. In particular,
\Eqn{SUSY} implies that at the SUSY scale, we have $\lambda_1 =
\lambda_2 = -(\lambda_3 + \lambda_4) = (g^2 + g_Y^2)/4$, and we can
naturally expect that at lower scales $\lambda_1$ and $-(\lambda_3 +
\lambda_4)$ will not deviate much from their boundary value, $(g^2 +
g_Y^2)/4$, while $\lambda_2$ can be expected to grow significantly.

\begin{figure}[h]
  \begin{center}
\includegraphics[scale=0.35]{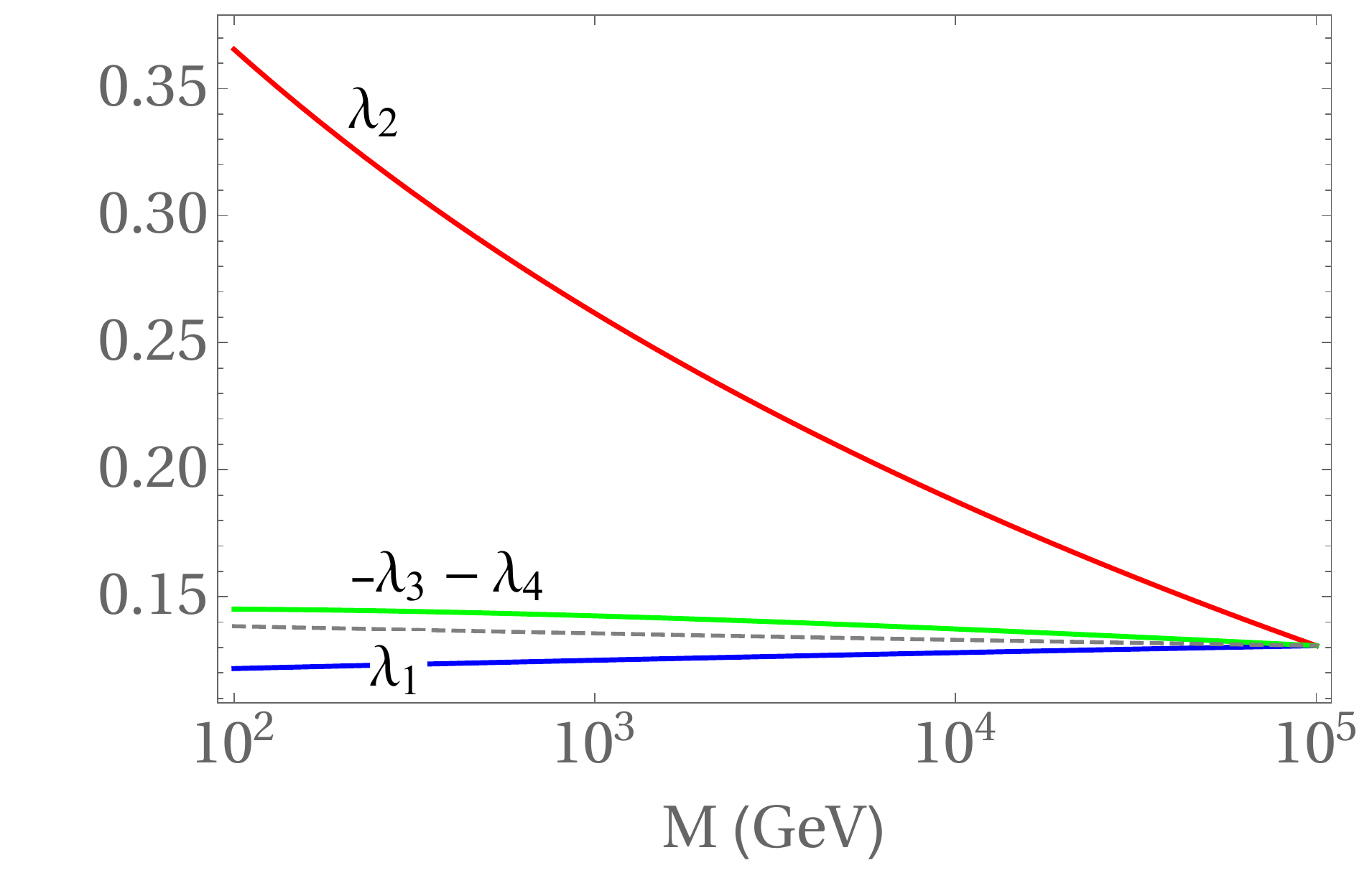} \quad
 \includegraphics[scale=0.35]{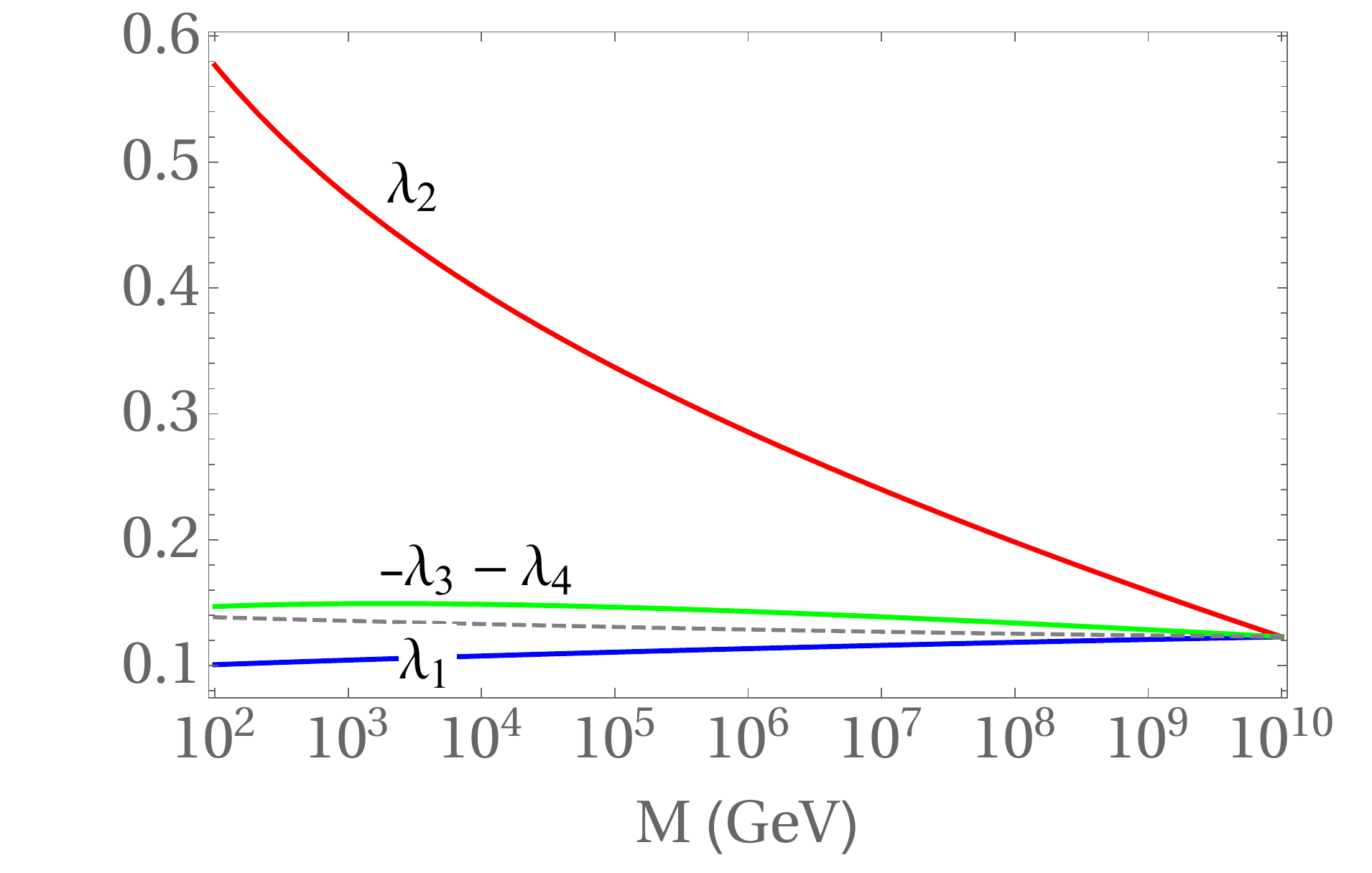}
\caption{\em Two-loop RG evolution of $\lambda_1$, $\lambda_2$ and
  $-(\lambda_3 + \lambda_4)$ starting from supersymmetric boundary
  values at $\Lambda_S = 10^5$~GeV with $\tan \beta=2.8$ (left panel),
  and $\Lambda_S = 10^{10}$~GeV with $\tan \beta=1.7$ (right panel),
  as compared to the evolution of $(g^2+g_Y^2)/4$ (dashed line).}
\label{f:levol}
\end{center}
\end{figure}
We can observe this behavior in Fig.~\ref{f:levol}, where we have used
two loop RGE to obtain the $\lambda_i$ values at the EWS starting
from supersymmetric boundary values at $\Lambda_S = 10^5$~GeV (left
panel) and $\Lambda_S = 10^{10}$~GeV (right panel).  Note that we used
the two loop RGE for the top quark Yukawa coupling because there is an
accidental cancellation in the one loop beta function for $\tan\beta
\sim 0.75$ which makes the two loop contributions relevant.  One can
already see this in the SM running of the top Yukawa coupling, ${\cal
  D}y_t^{\rm SM} \sim y_t^{\rm SM}\left\{4.5 (y_t^{\rm SM})^2 - 8
g_s^2\right\}$, which vanishes for $y_t^{\rm SM} = 4g_s/3$, where
$g_s$ is the gauge coupling for strong interaction. Strictly within
the SM framework, this numerical situation never arises. In the 2HDM,
however, the corresponding relation is $y_t^{\rm SM}/\sin\beta =
4g_s/3$.  This implies that in the vicinity of $\sin\beta \sim 0.6$,
{\it i.e.}, $\tan\beta \sim 0.75$, the one loop contributions to the
beta function can be overshadowed by the two loop ones.

\begin{figure}[h]
  \begin{center}
          \includegraphics[scale=0.35]{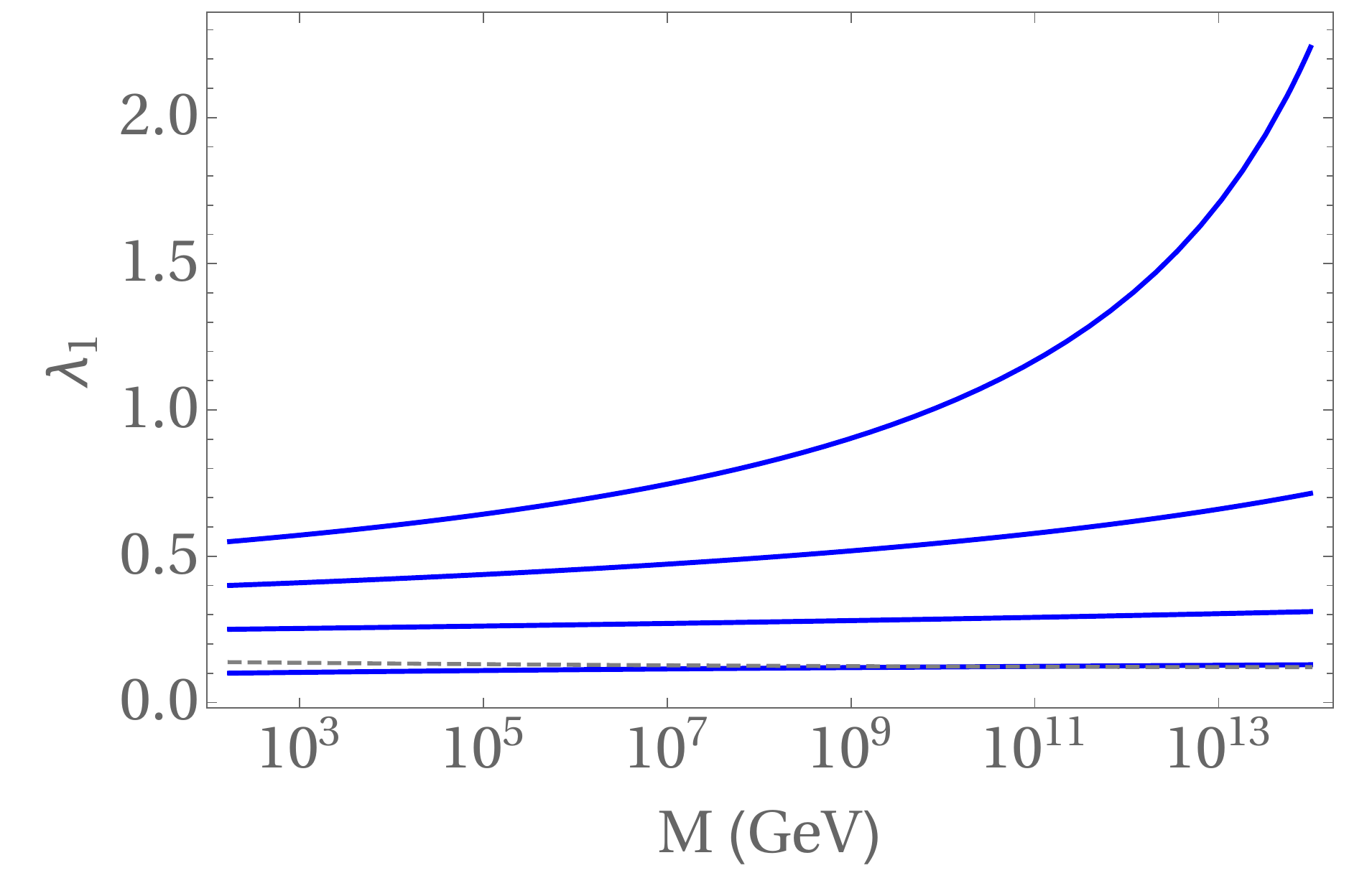}        
\caption{\em Evolution of $\lambda_1$ for different initial values (
  from bottom to top, $\lambda_1 = 0.10,0.25,0.40,0.55$), as compared
  with $(g^2+g_Y^2)/4$ (dashed line), for $\tan \beta=1.7$.}
\label{f:l1evol}
\end{center}
\end{figure}
In Fig.~\ref{f:levol} we showed that starting from small values
($\lambda_1\sim g^2,g^2_Y$) $\lambda_1$ does not run much from
$\Lambda_S$ to the EWS, and stays small. Now in a bottom-up approach
one may wonder if this is also true when starting with larger values
of $\lambda_1$ at the EWS and evolving it up to $\Lambda_S$. This
evolution is shown in Fig.~\ref{f:l1evol}. With respect to the other
quartic couplings, we take their initial EWS values to be: $\lambda_2 =
0.56$, $\lambda_3 = 0.015$ and $\lambda_4 = - 0.16$. Note, the
evolution of $\lambda_1$ is independent of $\lambda_2$ at one loop. With 
regards to $\lambda_3$ and $\lambda_4$, we take the relatively small
values corresponding to the gauge boundary conditions at the high
scale.  We see that, indeed, $\lambda_1$ evolves very little for small values of 
$\lambda_1$ at the EWS, $\lambda_1 \leq 0.40$, and this result is, in practice,
independent of $\tan \beta$ for $\tan \beta \leq 10$. Moreover
$\lambda_1$ grows with the scale, and therefore, at the EWS, we should
expect its value to be slightly smaller than $(g^2+g_Y^2)/4\simeq
0.15$, if it is indeed determined by gauge couplings at the high
scale.

From this discussion we can infer that a measurement of the quartic
couplings of the Higgs potential at the EWS can favor a high scale
SUSY scenario if the following features are observed:
\begin{itemize}
\item The values of $\lambda_1$ and $-(\lambda_3+\lambda_4)$, at the
  EWS, are in the vicinity of $(g^2 + g_Y^2)/4 \simeq 0.14$.
\item The value of $\lambda_2$ should then be significantly larger
  than $(g^2 + g_Y^2)/4$, due to the large negative contribution to
  the RGE from the top Yukawa coupling.
\item We can get a qualitative estimate of the SUSY scale,
  $\Lambda_S$, as the scale where $\lambda_2$ reaches its high scale
  boundary value, $(g^2 + g_Y^2)/4$.
\item If $\lambda_1$ (or $-(\lambda_3+\lambda_4)$) at the EWS is found
  to be larger than $\sim 0.4$, it will be impossible to satisfy the
  MSSM boundary conditions at a higher scale.
\end{itemize}

We emphasize that a generic 2HDM is not expected to have such
correlations among its quartic couplings. Therefore, the above
assertions would constitute a strong indication of a SUSY framework at a
high scale.

\begin{figure}[ht!]
	\centering
\includegraphics[width=5.2cm,height=4cm]{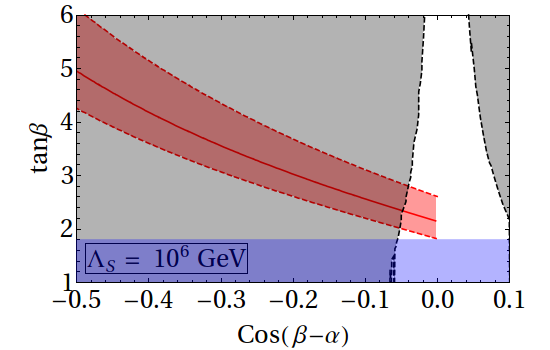}  
\includegraphics[width=5.2cm,height=4cm]{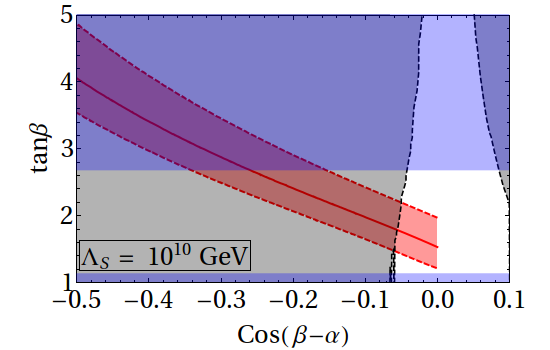}
\includegraphics[width=5.2cm,height=4cm]{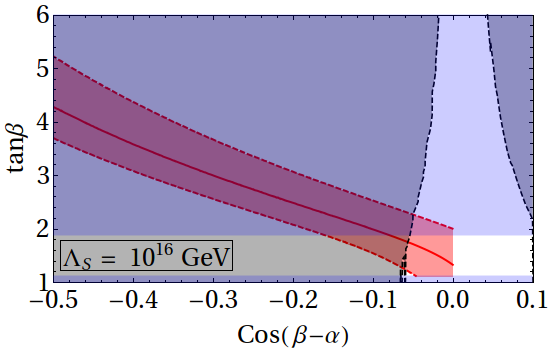} \\
\includegraphics[width=5.2cm,height=4cm]{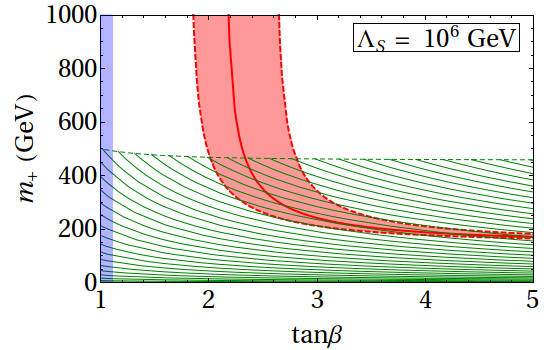} 
\includegraphics[width=5.2cm,height=4cm]{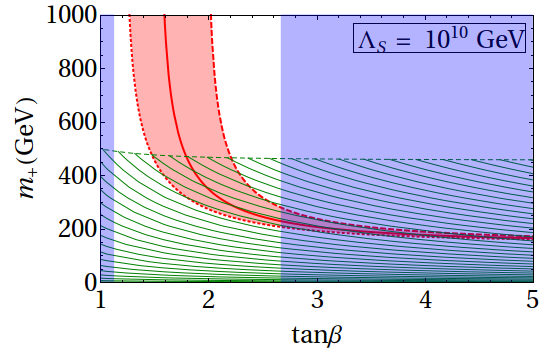} 
\includegraphics[width=5.2cm,height=4cm]{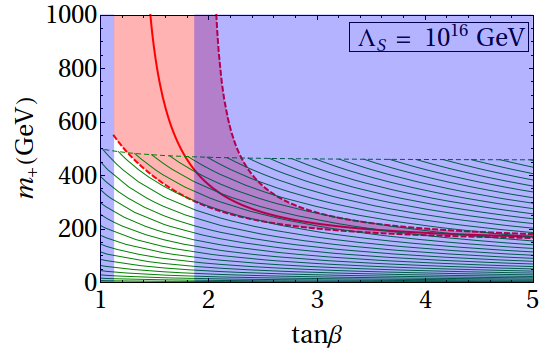} \\
\includegraphics[width=5.2cm,height=4cm]{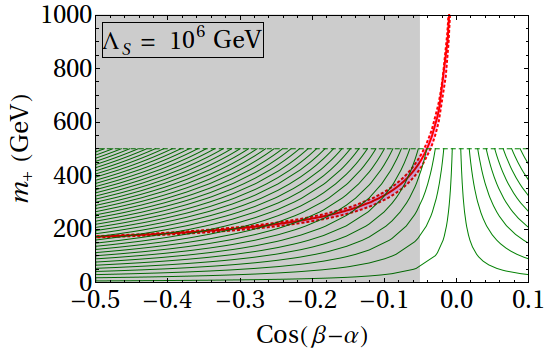} 
\includegraphics[width=5.2cm,height=4cm]{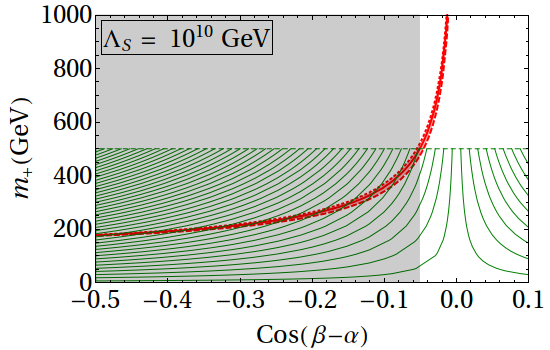} 
\includegraphics[width=5.2cm,height=4cm]{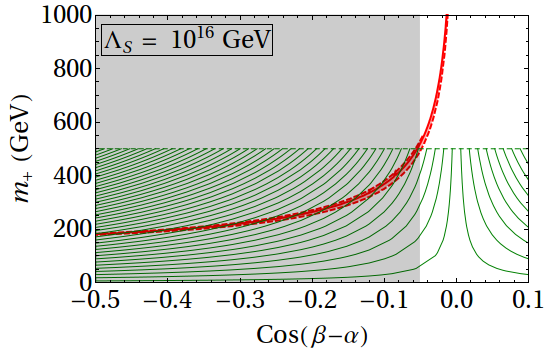} 
\caption{\em Solution curves in different planes for different choices
  of $\Lambda_S$. The widths of the solution regions (in red) arise
  from $2\sigma$ experimental uncertainties in $m_t$ and $m_h$. The
  regions disallowed from absolute stability (from $M_Z$ all the way
  to $\Lambda_S$) have been shaded in blue, while the hatched regions
  are disfavored from BR($b\to s\gamma$) at 95\% C.L. The shaded
  regions in gray are ruled out from the LHC Higgs data.}
\label{f:solutions}
\end{figure}

\section{Constraints and uncertainties of the SUSY scale determination} \label{sec:numerics}
To justify our choices for the EWS values of the quartic couplings and
$\tan\beta$ used in Fig.~\ref{f:levol}, we perform a numerical study
of the available parameter space at low energy, provided the quartic
couplings have been fixed by the supersymmetric boundary conditions at
 $\Lambda_S$.  We display our results in
Fig.~\ref{f:solutions} for three different choices of $\Lambda_S$.
Considering the fact that sub-TeV nonstandard scalars, for type~II
2HDM, are disfavored from flavor data for $\tan\beta < 1$
\cite{Deschamps:2009rh,Das:2015qva}\footnote{Although the flavor constraints
  mainly affect $m_+$, additional bounds from the $T$-parameter
  require $m_H, m_A$ and $m_+$ to be nearly degenerate
  \cite{Bhattacharyya:2013rya,Das:2015mwa}.}, we concentrate in the $\tan\beta >1$
region for possible interesting phenomenology.  The allowed parameter
region from this analysis, consistent with $m_h = 125.0 \pm 0.6 $~GeV
and a top pole mass $m_t = 173 \pm 1$~GeV, has been shaded in red. The
width of this region comes from the uncertainties in the input
parameters $m_t$ and $m_h$, around the central continuous line
corresponding to their central values. The values of $\tan \beta$
disfavored from absolute stability (from $M_Z$ to $\Lambda_S$) of the
scalar potential has been shaded in blue. The hatched region in the
middle and bottom panels of Fig.~\ref{f:solutions} is disfavored at
95\% C.L. from BR($b \to s\gamma$) \cite{Misiak:2017bgg}. The gray
shaded region in the top panel is forbidden by the Higgs data at 95\%
C.L.~\cite{Bernon:2014vta}.  The gray region in the bottom panel,
however, represents a disallowed region using a conservative bound on
$\cos(\beta-\alpha)$ from the Higgs data\cite{Bernon:2014vta}\footnote{
Although we have used the Run 1 data from the LHC to extract the bound
on $\cos(\beta-\alpha)$, the Run 2 data as summarized in 
Ref.~\cite{Chowdhury:2017aav}, does not significantly improve the limit.}.
Some of the interesting
features that emerge from Fig.~\ref{f:solutions} are summarized below.

\begin{wrapfigure}{r}{0.44\textwidth}
\centering
\includegraphics[scale=0.32]{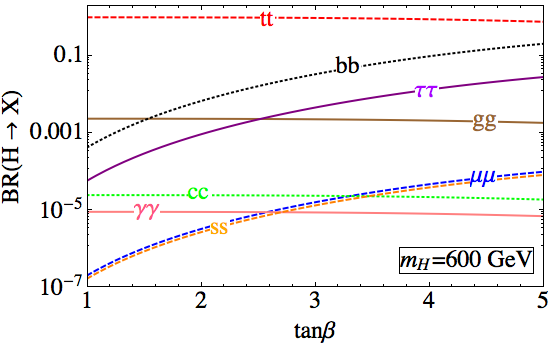}
\caption{\em Branching ratios of the heavier scalar, $H$ into
  different decay channels as a function of $\tb$. We have assumed
  $|\cba|\approx 0$ so that $H$ does not decay into a pair of gauge bosons.}
\label{f:Hdecay}
\end{wrapfigure}

\begin{enumerate}[$(a)$]

\item The main feature of Fig.~\ref{f:solutions}, as apparent from the
  top and middle panels, is that for a large supersymmetric scale,
  only low $\tan \beta$ values can reproduce the observed Higgs mass.
  Taking into account the constraints on $\cba$ and $m_+$ along with
  the requirement of absolute vacuum stability, we find that $1.8\leq
  \tan \beta \leq 2.8$ for $\Lambda_S = 10^6$~GeV, while $1.2\leq \tan
  \beta \leq 2.2$ for $\Lambda_S = 10^{10}$~GeV and $1.1\leq \tan
  \beta \leq 1.9$ for $\Lambda_S = 10^{16}$~GeV.  These results are in
  qualitative agreement with those in
  Refs.~\cite{Lee:2015uza,Bagnaschi:2015pwa} in the aspects where the
  analyses overlap.

\item It is interesting to note that for large $\Lambda_S$ we obtain
  an {\em upper} limit on $\tb$ from the requirement of absolute
  stability, in addition to a lower limit that stability usually
  offers in a generic 2HDM where the top Yukawa is proportional to
  $m_t/(v \sin\beta)$ \cite{Das:2015mwa}. This upper limit, which is
  rather strong ($\tb \lesssim 2$ for $\Lambda_S = 10^{16}$ GeV),
  arises from the requirement of satisfying $\lambda_3 + \lambda_4 +
  \sqrt{\lambda_1 \lambda_2} \ge |\lambda_5|$ \cite{Branco:2011iw} at
  all scales, but in our specific embedding of 2HDM in a SUSY
  backdrop.

\item From the $\cba$ vs $m_+$ plot, we see that we are practically in
  the decoupling region~\cite{Gunion:2002zf}. In the middle panel we
  observe that for a given value of $\tan \beta$, any value of $m_+
  \gtrsim 500$ GeV is possible when we take into account the
  uncertainties in the parameters\footnote{We have derived this limit
    from $b \to s \gamma$ admittedly from leading order contributions;
    a more recent analysis with higher order effects yields $m_+
    \gtrsim 570$ GeV \cite{Misiak:2017bgg}.}. As mentioned before, the
  allowed range of $\tan \beta$ depends on the scale, $\Lambda_S$. It
  is still, nevertheless, possible to have nonstandard scalars below
  the TeV scale, which is encouraging for the collider experiments.


\item As expected $m_+$ and $\cba$ are strongly correlated
  irrespective of the SUSY scale. This is easily understood as this
  mixing comes from the diagonalization of the neutral Higgs mass
  matrix in the Higgs basis, with offdiagonal elements ${\cal O}
  (v^2)$ and a large diagonal entry ${\cal O} (m_+^2)$.


\item As the allowed region in Fig.~\ref{f:solutions} is confined
  around $\tan\beta \sim \order(1)$ (for $\Lambda_s \geq 10^6$ GeV),
we have to focus on the left-most side of Fig.~\ref{f:Hdecay} 
where we display the decay pattern of the heavier CP even scalar $H$.

\end{enumerate}

As we discussed in Section \ref{sec:counting}, the variation of
$\lambda_2$ with the scale is more pronounced than that of other quartic
couplings, if we start from small boundary values at high
energy. Therefore, $\lambda_2$ is our best choice to determine
$\Lambda_S$ as the scale where it reaches its boundary value
$(g^2+g_Y^2)/4$. However, this evolution is very sensitive to the
values of $m_t$ and $\tan \beta$ at the EWS, as well as to the initial
$\lambda_2$ value.

\begin{figure}
  \begin{center}
          \includegraphics[scale=0.35]{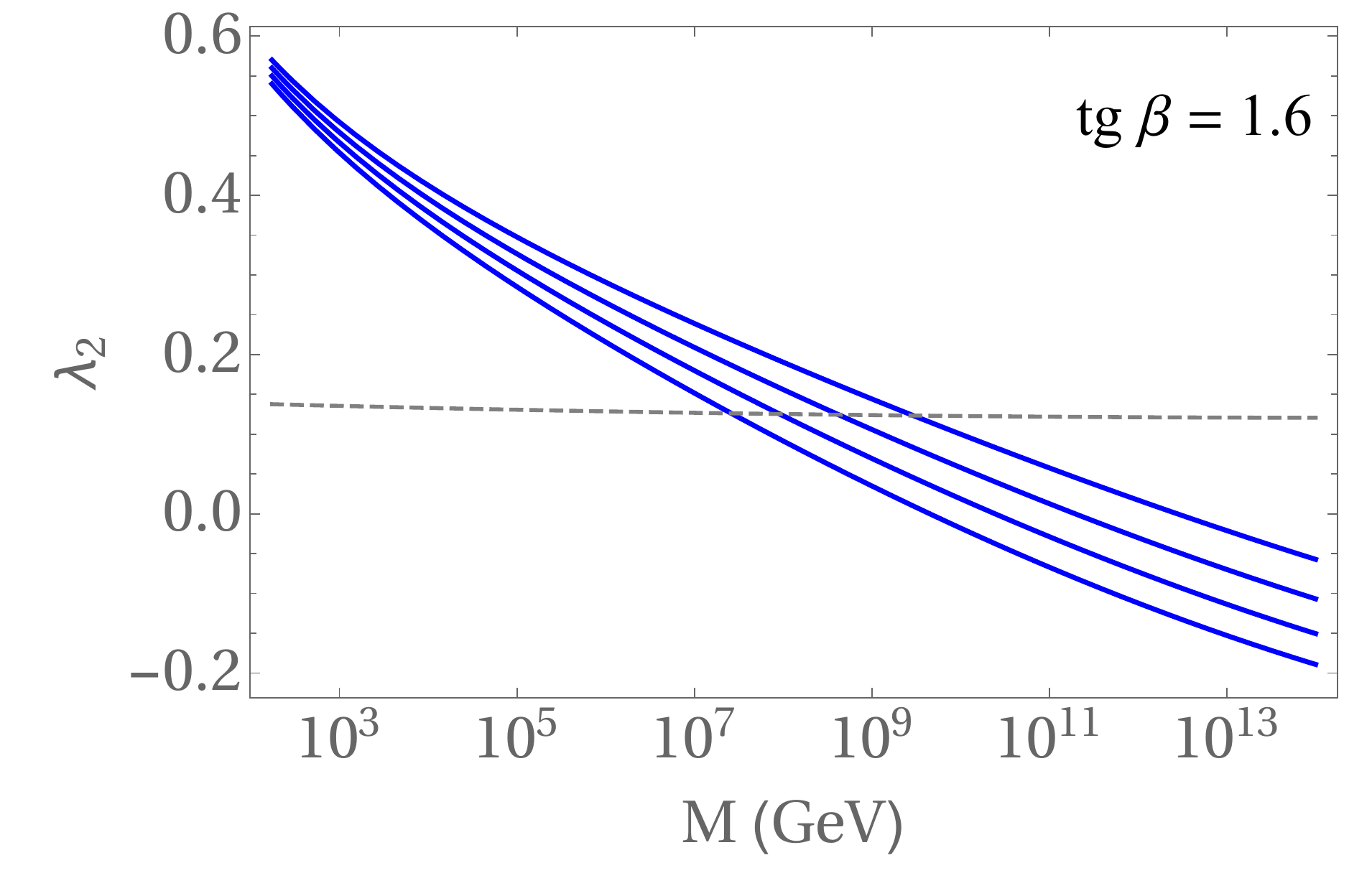}
          \includegraphics[scale=0.35]{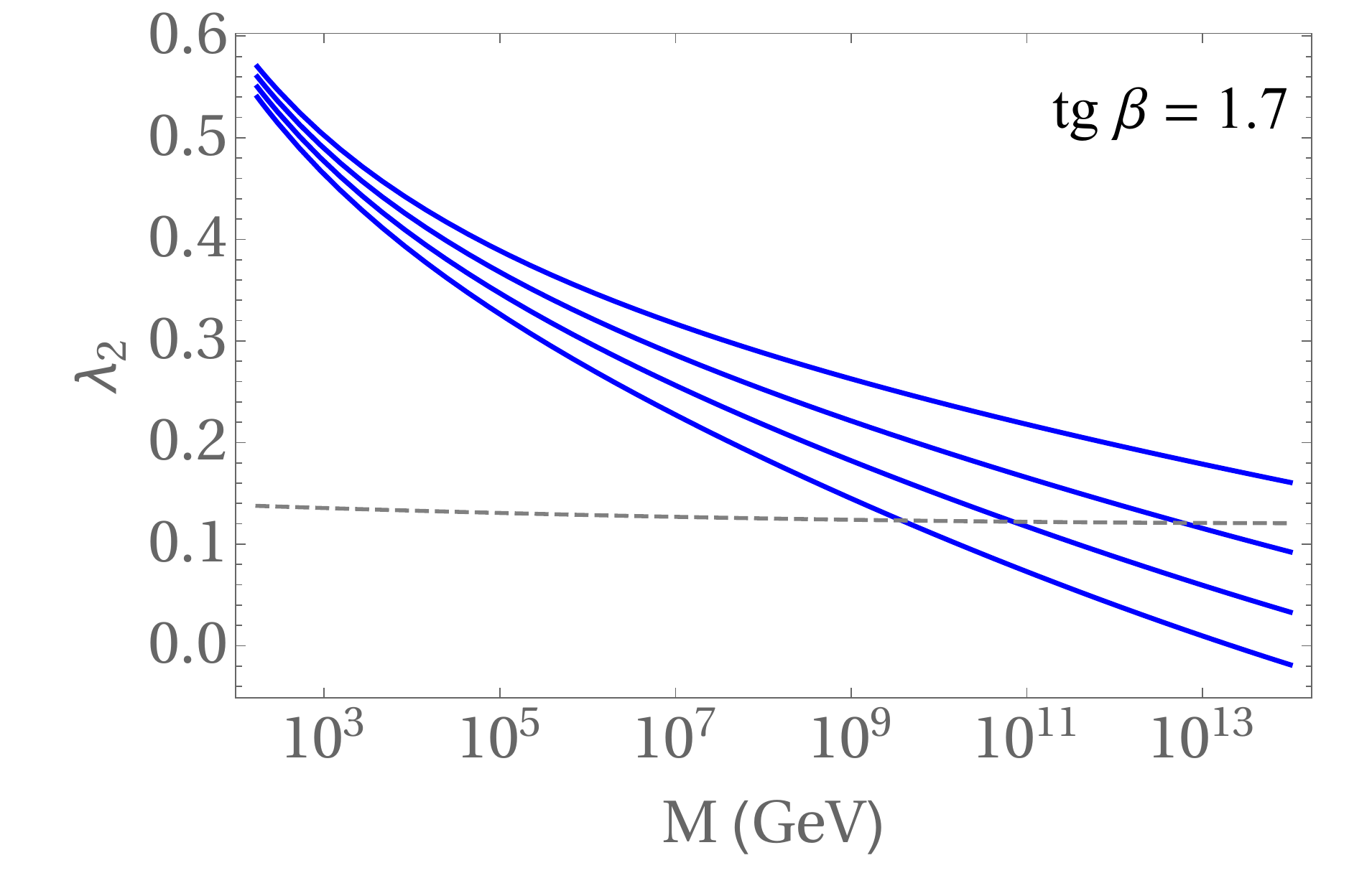}\\
          \includegraphics[scale=0.35]{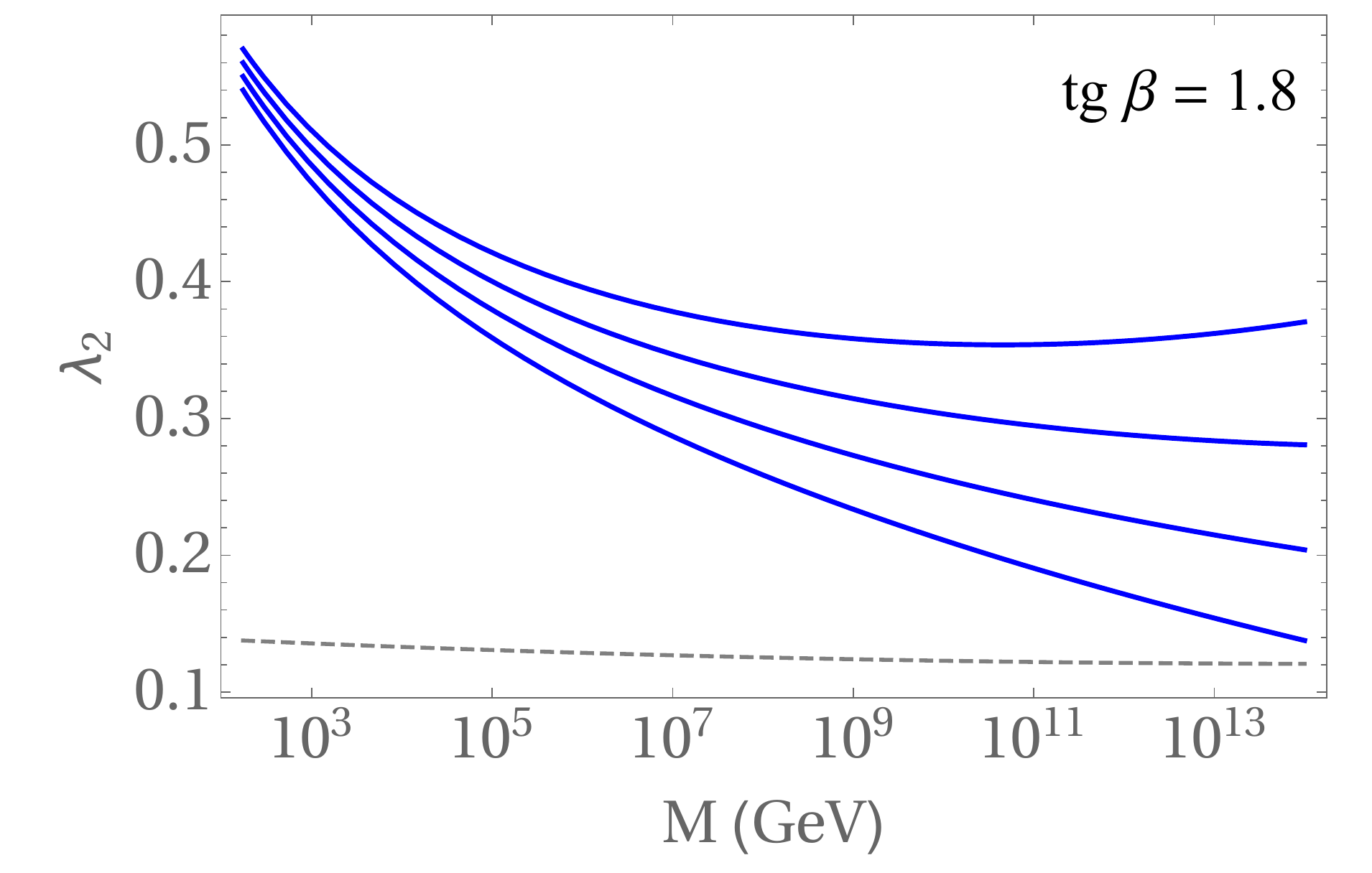}         
        \caption{\em Evolution of $\lambda_2$ (low to high scale) as a
          function of the scale $M$, for different initial values
          (from bottom to top, $\lambda_2 = 0.54,0.55,0.56,0.57$), as
          compared with $(g^2+g_Y^2)/4$ (dashed line) for different
          $\tan \beta$ values.}
        \label{f:l2evolu}
\end{center}
\end{figure}

This behavior of $\lambda_2$ is shown in Fig.~\ref{f:l2evolu}, where
we plot it for three similar values of $\tan \beta$ and several
closely spaced electroweak values of $\lambda_2$ consistent with the
observed Higgs mass. In fact, this figure is produced with a fixed
value of the top quark mass $m_t=173$ GeV; however, the intrinsic top
mass error of about $1$~GeV can be reproduced by a shift in
$\tan\beta$.  Given that the main effect of these uncertainties is a
change in the top Yukawa coupling, we can translate both uncertainties
as $\Delta \tan\beta=\tan\beta(1+\tan^2\beta)(\Delta
m_{t}/m_{t})$. From here, we obtain that $\Delta m_{t}= 1$~GeV
corresponds to $\Delta \tan\beta\sim 0.01$ for $\tan\beta=1$ and
$\Delta \tan\beta\sim0.06$ for $\tan\beta=2$. Thus, for the small
$\tan\beta$ values required when $\Lambda_S \geq 10^6$~GeV, the effect
of the top mass uncertainty is smaller than the effect of the $\tan
\beta$ range considered in Fig.~\ref{f:l2evolu}, but it grows as
$\tan^3\beta$ and will be important for larger values of $\tan\beta$.

Using Eqs.~(\ref{gij+}), (\ref{mHA+}), and (\ref{lambda1234}), we can
get an {\it a posteriori} explanation for the obtained values of
$\lambda_2$ and $\tan \beta$ at the EWS.  Under the assumptions of
sub-TeV nonstandard scalars and very small $\cba$, \Eqn{gij+} gives $g_{11}
v^2 \simeq m_h^2$. To a good approximation, we can also write
\begin{eqnarray}
\lambda_1(M_Z) \simeq \lambda_1(\Lambda_S) = \lambda_2(\Lambda_S)=
\frac{(g^2 + g_Y^2)}{4} = -\left\{\lambda_3(\Lambda_S)+
\lambda_4(\Lambda_S)\right\} \simeq -\left\{\lambda_3(M_Z)+
\lambda_4(M_Z)\right\} \,.
\label{equality:lambdas}
\end{eqnarray}
Now, using \Eqn{g11}, we obtain
\begin{eqnarray}
  m_h^2 = M_Z^2 \cos^2(2 \beta) + \Delta \lambda_2 v^2 ~\frac{\tan^4
    \beta}{\left( 1 + \tan^2\beta\right)^2} = M_Z^2 \left(\frac{\tan^2
    \beta -1}{\tan^2\beta +1}\right)^2 + \Delta \lambda_2 v^2 ~
  \left(\frac{\tan^2 \beta}{1 + \tan^2\beta}\right)^2\,,
  \label{mhsq}
  \end{eqnarray}
where, $\Delta \lambda_2 = \lambda_2(M_Z)-\lambda_2(\Lambda_S)$.
Eq.~(\ref{mhsq}) can easily be recognized as the usual expression for
the radiatively improved Higgs mass in the MSSM. This implies that the
mass of the observed Higgs boson is essentially determined by the RG
evolution of $\lambda_2$ and the value of $\tan \beta$. For a fixed
value of $\tan \beta$, the low energy value of $\lambda_2$ is uniquely
determined by $m_h$.  The larger the gap between $\Lambda_S$ and
$M_Z$, the more room $\lambda_2$ gets to grow under RG
evolution, thereby requiring a smaller $\tan\beta$ to reproduce the
observed Higgs mass.

To put our results into perspective, let us assume that all the
nonstandard scalar masses have been determined with an accuracy of 1
GeV {\it viz.,} $m_H = (503 \pm 1)$~GeV, $m_A=(491 \pm 1)$~GeV,
$m_+=(496 \pm 1 )$~GeV, and we have settled at $\cos (\beta - \alpha)
= -0.05$ and $\tan \beta= 1.7$.  These values would correspond to a
supersymmetric scale of $\Lambda_S \sim 10^{10}$ GeV.  However, it
should be noted that such an estimate of the SUSY scale is very
sensitive to the precise values of the input parameters, especially
$\tb$, and as shown in Fig.~\ref{f:l2evolu}; we would need to
determine $\tb$ at a few percent level to fix $\Lambda_S$
precisely. 
This ambiguity in the determination of the SUSY scale may partly be
	attributed to a common solution region for $\Lambda_S$ in the range
	of $10^6$-$10^{16}$~GeV, as apparent from 
	Fig.~\ref{f:solutions} (see point~(a) of Sec.~\ref{sec:numerics}
	also). On the other hand, if $\tan\beta$ turns out to be close to
	$2.2$ (say), then one can, for example, make a definitive conclusion
	that $\Lambda_S \le 10^{10}$~GeV.
Such a precise measurement of $\tan\beta$ would, perhaps, require us to wait for the
future linear colliders.  Nonetheless, the analysis presented in this
paper is good enough to provide an initial hint for the location of
the scale where SUSY is expected to appear.

\section{Conclusions} \label{sec:conclusions}
Our intention in this work was to address the question we posed in the
title as directly as possible.  To this end, we explored an effective 2HDM arising
from a more fundamental theory at a high scale, $\Lambda_S$, which
fixes the parameters of the Higgs potential. In particular, we have
focused on the high-scale MSSM as an example, where the Higgs quartic
couplings are determined by the supersymmetry breaking $D$-terms as functions
 of the gauge couplings. We have found that very high-scale MSSM
scenarios are still compatible with the observed Higgs mass for $\tb
\sim \order(1)$. Though our approach and emphasis is somewhat
different, we agree with the general conclusions of the existing
analyses studying effective 2HDMs stemming from high scale SUSY wherever we
overlap.

We emphasize that our methodology is quite general and can be applied
not only to SUSY but to a wide variety of UV scenarios in which all the quartic couplings
of the 2HDM potential of \Eqn{potential} are fixed at a high scale,
$\Lambda_S$.  For instance, we could have started with the assumption
that all the quartic couplings vanish at $\Lambda_S$.  For this
particular scenario, we find that the requirement of $m_h
\approx 125$ GeV, $v=246$ GeV and $|\cba| \sim 0$ and the absolute
stability of the potential up to $\Lambda_S$ favors a region of large
$\tb \sim 50$. In this region, the evolution of the quartic couplings
makes the charged scalar rather light, $m_+\approx 180$~GeV, which is
ruled out from the measurement of BR($b\to s\gamma$). This particular
scenario is, therefore, disfavored by the experimental data.

In the context of the supersymmetry, our analysis shows how possible (future) 
measurements of the nonstandard scalar masses, $\tb$ and $\cba$ can fix the
$\lambda_{1,2,3,4}$ couplings of the 2HDM potential, neglecting
$\lambda_{5,6,7}$, as is natural in the MSSM. Using the 2HDM
RGE we find that $\lambda_1$ and $-(\lambda_3+\lambda_4)$ should
stay close to their boundary value, $(g^2 + g_Y^2)/4$, all the way
from $\Lambda_S$ to $M_Z$, while $\lambda_2$ can grow significantly during 
RG running due to the large top Yukawa coupling.
This opens up the possibility of determining the supersymmetric scale, 
$\Lambda_S$, from the RG evolution of $\lambda_2$ as the scale where 
$\lambda_2$ reaches its boundary value, $(g^2 + g_Y^2)/4$. 
However, 
this strategy crucially depends on whether $\tb$ can be determined with a percent level precision in order to 
make a reasonable prediction for the MSSM scale;  a linear collider would be essential
to make further inroads.

\section*{Acknowledgements}
GB thanks the Physics Department of the University of Valencia (when
the work was initiated) and CERN Theory Division (during the final
stages of the work) for hospitality. GB acknowledges support of the
J.C.~Bose National Fellowship from the Department of Science and
Technology, Government of India (SERB Grant
No.\ SB/S2/JCB-062/2016). IS would like to thank 
Manuel Krauss for useful discussion.
This work is partially supported by the
Spanish MINECO under grants FPA2014-54459-P, FPA2017-84543-P by the Severo Ochoa
Excellence Program under grant SEV-2014-0398 and by the “Generalitat
Valenciana” under grants GVPROMETEOII2014-087 and PROMETEO-2017-033.


\bibliographystyle{JHEP}
\bibliography{references.bib}
\end{document}